\documentclass[sigconf]{acmart}

\AtBeginDocument{%
  }

\copyrightyear{2024}
\acmYear{2024}
\setcopyright{rightsretained}
\acmConference[ICCAD '24]{IEEE/ACM International Conference on Computer-Aided Design}{October 27--31, 2024}{New York, NY, USA}
\acmBooktitle{IEEE/ACM International Conference on Computer-Aided Design (ICCAD '24), October 27--31, 2024, New York, NY, USA}
\acmDOI{10.1145/3676536.3676840}
\acmISBN{979-8-4007-1077-3/24/10}

\usepackage{threeparttable}
\usepackage{enumitem}    
\usepackage{multirow}
\usepackage{fontawesome}
\usepackage{amsfonts}
\usepackage{pifont}
\newcommand{\cmark}{ \ding{51}}%
\newcommand{\xmark}{ \ding{55}}%

\newcommand{\SX}[1]{\textcolor{black}{#1}}

\begin{document}

\title[]{
Mixed-precision Neural Networks on RISC-V Cores: ISA extensions for Multi-Pumped Soft SIMD Operations}


\author{Giorgos Armeniakos}
\email{armeniakos@microlab.ntua.gr}
\orcid{XXXX-XXXX-XXXX}
\affiliation{%
  \institution{National Technical University of Athens}
  \city{Athens}
  \country{Greece}
  \postcode{XXXXX}
}

\author{Alexis Maras}
\email{amaras@microlab.ntua.gr}
\orcid{XXXX-XXXX-XXXX}
\affiliation{%
  \institution{National Technical University of Athens}
  \city{Athens}
  \country{Greece}
  \postcode{XXXXX}
}

\author{Sotirios Xydis}
\email{sxydis@microlab.ntua.gr}
\orcid{XXXX-XXXX-XXXX}
\affiliation{%
  \institution{National Technical University of Athens}
  \city{Athens}
  \country{Greece}
  \postcode{XXXXX}
}

\author{Dimitrios Soudris}
\email{dsoudris@microlab.ntua.gr}
\orcid{XXXX-XXXX-XXXX}
\affiliation{%
  \institution{National Technical University of Athens}
  \city{Athens}
  \country{Greece}
  \postcode{XXXXX}
}


\renewcommand{\shortauthors}{}

\newif\ifoutline
\outlinetrue

\newcommand{\red}[1]{{\color{black}#1}}
\newcommand{\blue}[1]{\ifoutline{\color{black}#1}\fi}
\newcommand{\orange}[1]{\ifoutline{\color{black}#1}\fi}

\begin{abstract}
Recent advancements in quantization and mixed-precision approaches offers substantial opportunities to improve the speed and energy efficiency of Neural Networks (NN). 
Research has shown that individual parameters with varying low precision, can attain accuracies comparable to full-precision counterparts.
\SX{However, modern embedded microprocessors provide very limited support for mixed-precision NNs regarding both Instruction Set Architecture (ISA) extensions and their hardware design for efficient execution of mixed-precision operations, i.e., introducing several performance bottlenecks due to numerous instructions for data packing and unpacking, arithmetic unit under-utilizations etc.}
In this work, we bring together, for the first time, ISA extensions tailored to mixed-precision hardware optimizations, targeting energy-efficient DNN inference on leading RISC-V CPU architectures.
We introduce a hardware-software co-design framework that supports cooperative hardware design, mixed-precision quantization, ISA extensions, and cycle-accurate emulations.
At the hardware level, we expand the ALU unit in our micro-architecture for configurable mixed-precision arithmetic operations and implement multi-pumping to reduce execution latency, with soft SIMD optimization for 2-bit operations.
At the ISA level, we encode three distinct MAC instructions extending the RISC-V ISA, each for different mixed-precision modes, and expose them to the compiler.
Our extensive experimental evaluation over widely used
DNNs and datasets, such as CIFAR10 and ImageNet, demonstrates that our framework can achieve, on average, 15$\times$ energy reduction for less than 1\% accuracy loss and outperforms the ISA-agnostic state-of-the-art RISC-V cores. 
\end{abstract}


\maketitle



\section{Introduction}


In the landscape of deep neural networks (DNNs), architectures are becoming increasingly complex, striving to maximize the accuracy of many modern applications, such as computer vision tasks, speech recognition and more~\cite{jouppi2017datacenter}.
In order to keep up with this increased computational complexity,  modern DNNs have widely adopted, as an integral part, quantization to reduce neural network's precision, i.e., the number of bits needed to load, store and process~\cite{Pulp-NN}.
Although traditionally NNs have relied on fixed precision formats to ensure accuracy and stability during training or inference, there is a rising need for different precision levels stemming from the observation that not all operations within a neural network require the same level of numerical precision.
Some layers or operations can tolerate lower precision without significant impact in quality, leading to a more efficient use of computational resources.

The concept of mixed-precision neural networks~\cite{Mix-GEMM} offers a promising solution, permitting variable precision across different network segments.
This adaptability can lead to considerable gains in performance and energy efficiency, since it allows for the maintenance of high precision where it is critical for accuracy, while reducing it in other areas to conserve computational costs~\cite{csurarme}.
Mixed-precision quantization can enable resource-constrained devices to support even larger models.
However, many existing general-purpose Central Processing Unit (CPU) architectures lack sufficient architectural support for efficiently managing fine-grained precision-mixing. 
Most Instruction Set Architectures (ISAs) either support coarse-grained schemes~\cite{XpulpNN} or offer sub-optimal formats~\cite{Eyeriss}, resulting in severe overheads during operand packing and unpacking.
Hence, advancing hardware-software co-design solutions and exploring different optimization layers of abstraction to harness the benefits of mixed-precision neural networks for energy-efficient data computation, while adhering to stringent area constraints, presents a significant challenge in computer architecture research.

In this work, we embrace the RISC-V ISA extension paradigm and enhance it with light-weighted hardware modifications in a small 32-bit RISC-V CPU core, Ibex~\cite{ibexproc}\footnote{\SX{We use Ibex core as our proof-of-concept microarchitecture. The concepts and the solution design can be straightforwardly adopted in other RISC-V cores.}}.
We propose an end-to-end automated framework that through design space exploration (DSE) generates and evaluates mixed-precision neural networks on modified RISC-V CPU processor.
To this end, we design, elaborate and integrate three innovative instructions that extend the RISC-V ISA and are portable on any modern RISC-V C compiler. 
These instructions have the capability to activate various operational modes within the modified microarchitecture, depending on the selected mixed-precision configuration.
At the microarchitecture level, we enable efficient operand packing and introduce soft SIMD instruction capabilities. Additionally, and at the circuit level, we effectively implement multi-pumping~\cite{10069694,multipump:2017} to fully leverage the enhanced operand bandwidth per mixed-precision instruction, thereby maximizing instruction throughput.
\newline
\textbf{Our main contributions in this work are the following:}

\begin{enumerate}[topsep=0pt,leftmargin=*]
    \item We introduce an automated framework\footnote{Our framework is available open-source at \url{https://github.com/alexmr09/Mixed-precision-Neural-Networks-on-RISC-V-Cores}} for generating mixed-precision DNNs and evaluating them through cycle-accurate RISC-V emulations.
    \item We extend RISC-V ISA with three novel instructions tailored to distinct hardware optimizations techniques and integrate our proposed architecture into a RISC-V based processor, used as a proof-of-concept.
    \item We explore the trade-offs between accuracy and speedup, alongside the introduced hardware overheads, and investigate the performance of four widely-used DNNs. Our findings reveal energy efficiency levels that surpass state-of-the-art approaches.
\end{enumerate}

More specifically, through extensive evaluation across popular image classification DNNs, on MNIST, CIFAR-10, and ImageNet datasets, we show that extending the ISA of a RISC-V Ibex processor with the proposed solution for efficient mixed-precision computations, it delivers an average energy efficiency of $0.9$ TOPS/W and up to $1.5$ TOPS/W for less than 1\% top-1 accuracy degradation due to model compression, \SX{i.e. achieving $10.9\times$ performance and $15.4\times$ energy gains, respectively, w.r.t. to the original Ibex architecture}. \SX{Interestingly, these gains are achieved with only a 26\% increase in area complexity}.
Additionally, we shown that our framework offers scalable performance based on a user's accuracy constraint, 
thus with lower quality criteria, the corresponding efficiency can increase up to $1.9$ TOPS/W (for up to $4.5$\% accuracy loss), outperforming significantly state-of-the-art approaches.
\section{background \& related work}

\begin{table}[t]
\setlength\tabcolsep{1pt}
\renewcommand{\arraystretch}{1.2}
\caption{Qualitative comparison of related works.}
\begin{threeparttable}
\begin{tabular}{c|ccccccccc|c}
\hline
\multicolumn{1}{l|}{} &
  \textbf{\cite{ottavi2020mixed}} &
  \textbf{\cite{XpulpNNV2}} &
  \textbf{\cite{lee2018unpu}} &
  \textbf{\cite{tcad20}} &
  \textbf{\cite{wang2024optimizing}} &
  \textbf{\cite{XpulpNN}} &
  \textbf{\cite{FILMQNNEF}} &
  \textbf{\cite{Mix-GEMM}} &
  \textbf{\cite{SySMOL}} &
  Ours \\ \hline
Mixed Prec.     & \textcolor{teal}{\cmark} & \textcolor{teal}{\cmark} & \textcolor{teal}{\cmark} & \xmark & \xmark & \textcolor{teal}{\cmark} & \textcolor{teal}{\cmark} & \textcolor{teal}{\cmark} & \textcolor{teal}{\cmark} & \textcolor{teal}{\cmark} \\
ASIP\tnote{1}            & \textcolor{teal}{\cmark} & \textcolor{teal}{\cmark} & \xmark & \xmark & \textcolor{teal}{\cmark} & \textcolor{teal}{\cmark} & \xmark & \textcolor{teal}{\cmark} & \textcolor{teal}{\cmark} & \textcolor{teal}{\cmark} \\
Fine-grained    & \xmark & \xmark & \xmark & \xmark & \xmark & \xmark & \textcolor{teal}{\cmark} & \xmark & \textcolor{teal}{\cmark} & \textcolor{teal}{\cmark} \\
HW-aware train. & \xmark & \xmark & \xmark & \xmark & \xmark & \xmark & \textcolor{teal}{\cmark} & \textcolor{teal}{\cmark} & \textcolor{teal}{\cmark} & \textcolor{teal}{\cmark} \\
HW Optimiz.         & \textcolor{teal}{\cmark} & \xmark & \textcolor{teal}{\cmark} & \textcolor{teal}{\cmark} & \textcolor{teal}{\cmark} & \textcolor{teal}{\cmark} & \textcolor{teal}{\cmark} & \textcolor{teal}{\cmark} & \textcolor{teal}{\cmark} & \textcolor{teal}{\cmark} \\
Multipumping    & \xmark & \xmark & \xmark & \xmark & \xmark & \xmark & \xmark & \xmark & \xmark & \textcolor{teal}{\cmark} \\
Packing         & \xmark & \xmark & \xmark & \xmark & \xmark & \xmark & \textcolor{teal}{\cmark} & \xmark & \xmark & \textcolor{teal}{\cmark} \\ \hline
\end{tabular}
\begin{tablenotes}\small
\item[] $^1$Application-Specific Instruction set Processor 
\vspace{-2ex}
\end{tablenotes}
\end{threeparttable}
\label{tab:related}\vspace{-2ex}
\end{table}

In recent years, numerous studies have explored various compression optimization techniques, such as pruning and quantization~\cite{zervtc}, to minimize the inference latency and memory footprint of DNNs on edge devices. Optimized software libraries such as ARM's CMSIS-NN~\cite{CMSIS-NN}, Google's GEMMLowp~\cite{jacob2022gemmlowp}, and X-CUBE-AI from STMicroelectronics~\cite{STM-XCUBE-AI}, along with advanced frameworks like DORY~\cite{Dory} for RISC-V PULP platform processors and MCUNet~\cite{MCUNet,CMix-NN}, are widely used to deploy Quantized Neural Networks on commercial devices. Nonetheless, these tools primarily support 8-bit or higher precision variables and cannot fully exploit the computational efficiencies offered by lower-bit quantization of DNNs.

Since most modern systems are unable to efficiently compute sub-byte calculations, many works have focused on developing specialized DNN accelerators~\cite{YodaNN,tcad20,Eyeriss,Barvinn,DNN_Intel_FPGA}. These accelerators often provide significant performance enhancements, but those benefits are often compromised due to their lack of scalability, extensive area occupation, and high power consumption, typically in the range of hundreds of milliwatts~\cite{YodaNN,tcad20,Eyeriss} to a few watts~\cite{Barvinn}.

Alternatively, other works have looked into extending the RISC-V ISA and developing custom functional units specifically designed to handle efficient sub-byte operations. PULP-NN~\cite{Pulp-NN} achieves significant speed improvements by enabling lower-bit DNN inference, utilizing instructions that pack and extract vectors of smaller data sizes, such as 4-bit and 2-bit. However, these specialized casting instructions introduce overheads for computations at 4-bit and 2-bit sizes, diminishing the performance benefits associated with these lower bit-widths. XpulpNN~\cite{XpulpNN} supports 2, 4, and 8-bit SIMD operations but does not support mixed-precision computations.

Recent studies~\cite{ottavi2020mixed,lee2018unpu,Mix-GEMM,SySMOL} examined the utilization of mixed precision variables on resource-constrained devices as an effective strategy to balance the trade-offs between accuracy, latency, and memory footprint~\cite{wang2024optimizing} that arise when employing lower precision numbers. UNPU~\cite{lee2018unpu} explored bit-serial MAC units with consistent activation data size and weight data sizes spanning 1 to 16 bits, while~\cite{ottavi2020mixed} extended a RISC-V core by incorporating 4- and 2-bit MAC units alongside customized controllers to enable 2-/4-/8-bit mixed-precision computations. In Mix-GEMM~\cite{Mix-GEMM} and SySMOL~\cite{SySMOL}, the authors presented software/hardware co-design workflows that seamlessly integrate hardware accelerators into the processor architecture complemented by hardware-aware training. However, in~\cite{Mix-GEMM} they only consider per-network quantization in their evaluation, potentially overlooking the benefits of more granular precision adjustments, while~\cite{SySMOL} restricts its use to 1, 2, and 4-bit variables within intra-layer quantization, which can lead to significant accuracy losses in more complex models or challenging datasets like ImageNet. Furthermore, neither study explores low-level hardware optimizations, such as overclocking~\cite{Approximate_Computing_Survey} or packing multiple low-precision multiplications on a single block~\cite{FILMQNNEF} to optimize their system’s throughput.

Table~\ref{tab:related} summarizes the above discussion and provides a qualitative comparison of most relevant works in the field of RISC-V ISA extension for mixed-precision DNNs.
Our work distinguishes from most state-of-the-art approaches, since, to the best of our knowledge, none has proposed mixed-precision instructions capable of simultaneously activating different computations and operational modes through logic and circuit-level optimizations, such as multi-pumping and soft SIMD techniques.
In Section~\ref{sec:soacomp} we evaluate our work against most relevant state-of-the-art approaches.


\section{Configurable mixed-precision architecture based on Ibex core}

\begin{figure}[ht]
    \centering
    \includegraphics{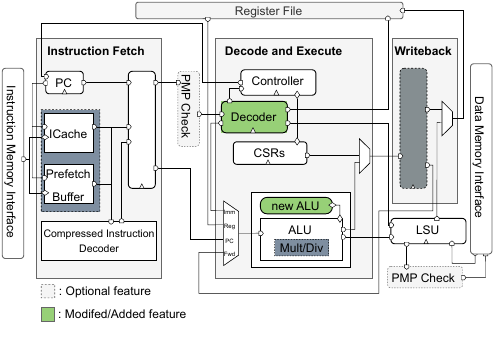}
    \caption{Ibex core architecture showcasing our modified/added features.}
    \label{fig:ibex}\vspace{-2ex}
\end{figure}

The base design utilized in this work is a generic microarchitecture implementation of the Ibex~\cite{ibexproc}, an open-source 32 bit RISC-V CPU core (Fig.~\ref{fig:ibex}).
Without loss of generality, the specific RISC-V core forms a proof-of-concept micro-architecture scenario to showcase the impact when enabling in an execution unit the support of fined-grained mixed-precision operations and can be also generalized in a straightforward manner to other RISC-V CPU cores.

As shown in Fig.~\ref{fig:ibex}, in Ibex, the instruction fetch stage retrieves instructions from memory, which are then passed to the decode and execution stage. 
Here, the instructions are decoded to determine their operation and operands, and the ALU performs arithmetic/logic operations accordingly. 
Finally, the results are written back to the register file in the writeback stage, completing the instruction cycle and enabling subsequent instructions to be executed.
The decode logic also needs to be augmented to accommodate any newly added instructions and to signal the ALU or the added Unit about the specific operation to be performed.
All other processor logic, such as controlling, forwarding logic, load/store units, etc., remains unchanged.

\subsection{Configurable Mixed-Precision Unit Design }


In this section, we detail our approach to extending the RISC-V CPU core to support fine-grained mixed-precision operations, coupled with hardware optimizations, and thereby improving run-time/energy efficiency, albeit with a small, but yet acceptable, degradation in accuracy.
The key idea is to allow multiple parallel operations (i.e., MAC/MUL units) by leveraging existing logic in architectures such as multiplication units, thereby introducing minimal hardware overheads.

\begin{figure}
    \centering
    \includegraphics{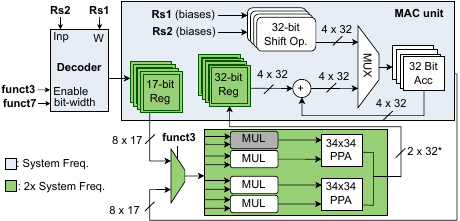}
    \caption{Microarchitecture of the modified ALU featuring packed operations, multi-pumped units and soft SIMD optimizations. Additional multiplier is highlighted in gray. (PPA: Partial Product Addition)}
    \label{fig:micro}
\end{figure}

To achieve this, we propose and implement a modified version of Ibex's generic multiplier, expanding its capability to handle variable mixed-precision operations.
Since we primarily focus on further optimizing the performance, rather than reducing total area, we firstly select as our baseline unit the one-cycle multiplier (RV32M)~\cite{ibexproc}, featuring three parallel 17-bit $\times$ 17-bit multiplication units and a 34-bit accumulator (\ref{fig:micro}).
Then, we expand this unit by integrating an additional 17-bit multiplier (gray MUL in Fig.~\ref{fig:micro}), allowing for parallel execution of finer-grained mixed-precision multiplications with varying bit-widths.
Interestingly, \SX{in Ibex's FPGA prototype}, the additional 17-bit multiplier accommodates the existing DSPs (i.e., 4 in total) and thus introduces minimal overhead.

Our primary objective is to minimize the total required cycles of the core operation of convolution and full-connected layer, which is given by:

\begin{equation}\label{eq:convar}
    Y_o = b_o + \sum_{i=1}^{N} W_{o,i}A_i,
\end{equation}
where $W_{o,i}$ are the filter’s weights, $A_i$ are the input activations, and $N$ is the number of weights.
Hence, we focus on the partial product computation of $W_{o,i}A_i$ and we consider the precision of all inputs to 8-bit, i.e., the smallest precision in
which accuracy is close to floating-point one for all models \red{(see Section~\ref{sec:evalframe})}, while we refine the precision of weights to either 2,4 or 8-bit.
The latter different configuration lead to three different modes, where for each mode we apply additional optimization techniques.

\subsection{\SX{Mixed-precision Micro-architectural Extensions}}
\begin{figure*}[ht]
    \centering
    \includegraphics[width=\textwidth,height=65mm]{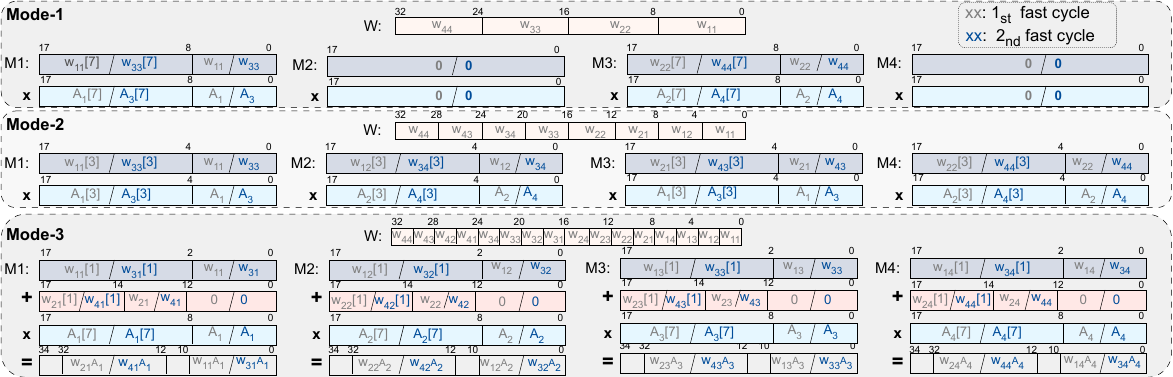}
    \caption{Illustrative example showing MAC operation optimization and mapping onto four multipliers (M) for three different mixed-precision configurations/modes. Only a partial load of weights and inputs is depicted for simplicity.}
    \label{fig:modes}\vspace{-2ex}
\end{figure*}

\textbf{Decoder extensions for Mixed-precision:} 
\red{
Since the initial architecture does not have the precision controllability, we modify the existing decoder (see Fig.~\ref{fig:ibex}), including extra control logic and multiplexers. This enhanced decoder facilitates the division of low-precision operands (i.e., activations and weights) and issues the appropriate control signals to the other blocks, allowing them to correctly execute the instruction corresponding to a specific operational mode.

\textbf{Mixed precision Arithmetic Logic Unit Design:}
Then, to leverage the full potential of different low precision configurations, we employ at different layers of abstraction distinct optimization strategies.
At the algorithmic level, and since the bit-width of weights is either 2,4 or 8-bit, the initial step involves packing (up to 16) operands (weights) into 32-bit registers.
}
\red{Each of the resulting pairs of weights and activations is mapped onto a single 17-bit $\times$ 17-bit multiplier within our modified unit.}
This procedure is repeated for the remaining three multipliers.
The latter approach, not only allows us to boost the parallelization of operations, but it also decreases the number of instructions required by initial (non-scaled tactic) loads and stores. 
Then, motivated by the fact that smaller parallel computing units can achieve higher maximum clock frequencies, we employ at circuit level a multi-pumping technique.
Typically, multi-pumping is used to minimize resource usage, rather than execution latency~\cite{multipump:2017}.
However, in this work we utilize multi-pumped units to accelerate the processing of packed operands within the core's pipeline, ensuring a flow without stalls and thereby reducing the total execution cycles.
Hence, we implement a multi-pumping scheme with 2x the clock frequency for all low precision MAC operations.
Lastly, and specifically only when 2-bit operands are used for weights, we apply a soft SIMD technique, making full use of the 17-bit multiplier's resources and handling as many MAC operations as possible.
To ensure that the output remains correct when multiple multiplications are packaged together, guard bits are placed between two inputs.
An example of this mode within a multiplier (e.g., M1) can be shown in Fig.~\ref{fig:modes}c (Mode-3), where we assume 2-bit for weights and 8-bit for activations:
\begin{equation}
    (A_1)\cdot(W_{21}\cdot2^{11} + W_{11}) = A_1W_{21}\cdot2^{11} + A_1W_{11}
\end{equation}
The first product $W_{11}A_1$ is placed on the 10 LSBs, in their respective ports.
The next product $W_{21}A_1$ need to be shifted at least 10-bits for correct calculations, while the 2-bit guard is for maximizing the use of each multiplier's resources.

Summarizing, our three different mixed-precision configurations with the aforementioned optimization techniques, can be categorized into the following modes:
\begin{enumerate}[topsep=0pt,leftmargin=*]
\item $Mode\!-\!1$ (low-speed): 
if 8-bit precision is used for weights (Fig.~\ref{fig:modes}a), then multi-pumping has zero impact, since only 4 weights are loaded in one 32-bit register during the first iteration. 
Speedup is accomplished only due to multipliers parallelization.
\item $Mode\!-\!2$ (medium-speed): if 4-bit precision is used for weights (Fig.~\ref{fig:modes}b), both multi-pumping and parallelization contribute to processing acceleration.
\item $Mode\!-\!3$ (high-speed): if 2-bit precision is used for weights (Fig.~\ref{fig:modes}c), our third INT2 optimization technique is also applied. In other words, $Mode\!-\!3$ is an optimized derivative of $Mode\!-\!2$.
\end{enumerate}


\begin{figure}
    \centering
    \includegraphics{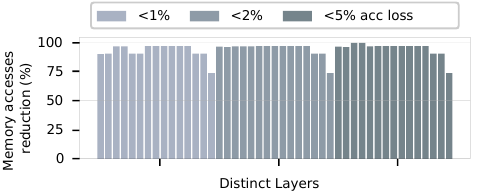}
    \caption{Memory accesses reduction, compared to the original Ibex architecture, delivered by our new instructions when mapping mixed-precision operation . Distinct layers of 3 mixed-quantized MobileNetV1 models were examined.}
    \label{fig:memory}\vspace{-3ex}
\end{figure}

\subsection{RISC-V ISA Extension}

To illustrate the speedup of our modes, we extend the RISC-V ISA to support three different operations, following the RISC-V ISA manual for custom extensions. These three instructions use the R-type format with six sub-fields: opcode, func3, func7, rs1, rs2, and rd, all 32 bits long. The opcode specifies the operation and the involved source (rs1, rs2) and destination (rd) registers. Func3 and func7 define the operations based on the input format.

Our RISC-V mixed-precision extension instructions are listed in Table~\ref{tab:isa}. We introduce only a few instructions, having no impact on the CPU's opcode decoder hardware. Specifically, $nn/_mac/_8b$ corresponds to $Mode-1$ and requires four 8-bit packed weights for four parallel MAC operations. $nn/_mac/_4b$ and $nn/_mac/_2b$ correspond to $Mode-2$ and $Mode-3$, requiring eight 4-bit and 16 2-bit packed weights, respectively. In all cases, the accumulator (rd) length is 32 bits. After completing the MAC operations for each output feature, the results are fetched from the 32-bit accumulators, followed by a common requantization step to adjust the output bit-width back to 8 bits.

The key takeaways of our instructions from enabling different operational $Modes$ are twofold.
Firstly, they facilitate a higher number of MAC operations per cycle, leading to increased throughput. 
Secondly, they facilitate considerably fewer loads and stores, leading to a substantial reduction in memory accesses. 
For example, the memory access reduction per layer for MobileNetV1 is depicted in Fig.\ref{fig:memory}.
To generate Fig.\ref{fig:memory}, we examined three mixed-precision models, ranging from less aggressive (<1\% accuracy loss) to more aggressive ones (up to 5\%).
This systematic exploration is detailed in Section~\ref{sec:framework}.
As illustrated, even with minimal quality degradation, in addition to accelerated MAC processing, memory accesses across different layers are reduced by an average of 85\%.

\textbf{Compiler support:} Once the encoding of instructions has been done, we provide a high-level interface for their utilization. 
Our approach involves implementing a C intrinsic for each instruction, utilizing the inline assembly $\_\_asm\_\_$ operator to emit the bytecode corresponding to the specific instruction. 
This approach avoids us to integrate code generation within the compiler.
To facilitate the inline assembly support for our custom instructions, we made minor adjustments to the RISC-V GNU toolchain within GCC's binutils. This modification enables the execution of any RISC-V binary on various architectures (e.g., x86 of a host machine). 
Finally, the compiled binaries, along with the complete instruction set containing all our extensions, are executed and thoroughly emulated using the Spike simulator~\cite{spike}.

\begin{table}[h]
\setlength\tabcolsep{1.2pt}
\renewcommand{\arraystretch}{1.1}
\footnotesize
\caption{Overview of the mixed-precision ISA extension encoding.}
\begin{tabular}{|c|c|c|c|c|c|c|}
\hline

 &
  \textbf{func7} &
  \textbf{func3} &
  \textbf{rs1} &
  \textbf{rs2} &
  \textbf{rd} &
  \textbf{Description} \\ \hline
$nn\_mac\_8b$ &
  000 1000 &
  010 &
  \begin{tabular}[c]{@{}c@{}}4 8-bit \\ activations\end{tabular} &
  \begin{tabular}[c]{@{}c@{}}4 8-bit\\ weights\end{tabular} &
  \begin{tabular}[c]{@{}c@{}}32-bit\\ Acc\end{tabular} &
  \begin{tabular}[c]{@{}c@{}}4 parallel MAC\\ (Mode-1)\end{tabular} \\ \hline
$nn\_mac\_4b$ &
  000 0100 &
  010 &
  \begin{tabular}[c]{@{}c@{}}4 8-bit \\ activations\end{tabular} &
  \begin{tabular}[c]{@{}c@{}}8 4-bit\\ weights\end{tabular} &
  \begin{tabular}[c]{@{}c@{}}32-bit\\ Acc\end{tabular} &
  \begin{tabular}[c]{@{}c@{}}8 parallel MAC\\ (Mode-2)\end{tabular} \\ \hline
$nn\_mac\_2b$ &
  000 0010 &
  010 &
  \begin{tabular}[c]{@{}c@{}}4 8-bit \\ activations\end{tabular} &
  \begin{tabular}[c]{@{}c@{}}16 2-bit\\ weights\end{tabular} &
  \begin{tabular}[c]{@{}c@{}}32-bit\\ Acc\end{tabular} &
  \begin{tabular}[c]{@{}c@{}}16 parallel MAC\\ (Mode-3)\end{tabular} \\ \hline
\end{tabular}
\label{tab:isa}
\end{table}
\section{proposed co-design framework}\label{sec:framework}

    
    

\begin{figure}[ht]
    \centering
    \includegraphics{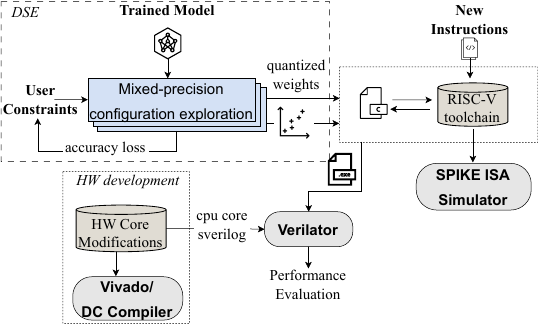}
    \caption{Workflow of our co-design framework.}
    \label{fig:workflow}
\end{figure}

In this Section, we present the workflow of our automated framework (Fig.~\ref{fig:workflow}) for generating mixed-precision neural networks and evaluating them on RISC-V CPU cores.
Briefly, our framework receives as an input a trained model (e.g., dumped from Pytorch) and performs a systematic design space exploration (DSE) to reveal mixed-precision configurations, offering different trade-offs (considering also user's desired accuracy threshold) among run-time efficiency and network accuracy. 
Once the functionality of the new instructions has been simulated and verified, the linker combine them into a single executable file (ELF).
The later, along with the RTL descriptions of the modified core, which are composed in System Verilog and synthesized using industrial strenglth EDA tools (Xilinx Vivado for FPGA flow and Synopsys Design Compiler for ASIC flow), are received as inputs from the Verilator~\cite{verilator}, a cycle-accurate emulator. 
Subsequently, performance metrics considering various accuracy-speedup trade-offs and any introduced overheads can be assessed.

\textbf{Mixed-precision exploration:} 
Our mixed-precision configuration design space exploration (DSE) starts with post-training quantization, where we systematically adjust the precision of various layers within the trained neural network.
To retain the accuracy loss, especially in complex datasets, we also perform a fine-tuning process with few extra epochs. 
Without loss of generality, we explore (but not limit) bit-width configurations of 2-bit, 4-bit, and 8-bit for both convolutional and fully-connected layers, which are the most computationally intensive layers~\cite{zervtc}.
To comprehensively assess trade-offs, we test all possible combinations of precision settings across these layers.
Note that this exploration needs to be performed offline and only once.
The total number of configurations directly depends on the number of layers $L$ and on the number of considered precisions $p$. 
Thus, the number of possible mixed-precision configurations is given by: $p^L$. 
While this number is manageable for small networks, it becomes exceedingly large for larger DNNs.
Consequently, particularly for large networks and based on experimental observations, we have chosen to strategically prune the design space, by setting a fixed high precision, i.e., 8-bit, for the sensitive initial layers, thereby simplifying the exploration process and focusing on where precision changes yield the most significant impact.
Hence, we manage to decrease on average more than $2000\times$ explored configurations, while still retaining numerous options that ensure high accuracy.
Indicatively, in our worst case scenario, for the MobileNetV1 architecture, where 1408 different configurations were examined, the model was retrained for an additional 35 epochs and our framework required less than 15hours, referring to an NVDIA T4 GPU and 16GB RAM. 

Given a selected mixed-precision configuration among the obtained Pareto space, our framework evaluates the model as follows:
\begin{enumerate}[leftmargin=*]
    \item Read the C source code that includes the respective replacements of the original kernels with kernels incorporating the \textit{nn\_mac\_(x)b} operations.\label{item:s1}
    \item Update the RISC-V toolchain with the new instructions, verify their functionality in Spike simulator and generate the ELF executable file.\label{item:s2}
    \item Synthesize the modified core using EDA tools for both FPGA and ASIC technology, and extract the RTL descriptions.\label{item:s3}
    \item \red{Simulate the ISA-extended Ibex processor using Verilator and along with Synopsys PrimeTime extract performance metrics and power analysis.}\label{item:s4} 
\end{enumerate}
Note that only step~\ref{item:s1} and~\ref{item:s4} need to be executed for different benchmarks or configurations, while steps~\ref{item:s2} and~\ref{item:s3} are executed only once.
Finally, our framework can accommodate all types of neural networks, as long as their layers conform to the generalized parallel neural MAC operations.

\section{Evaluation \& Results}


\subsection{Experimental Setup}

\begin{table}[t]
\setlength\tabcolsep{5pt}
\renewcommand{\arraystretch}{1.1}
\small
\caption{Description of our evaluated baseline models}
\begin{threeparttable}
\begin{tabular}{|c|c|c|c|c|}
\hline
\textbf{Model} & \textbf{Acc (\%)} & \textbf{Topology\tnote{1}} & \textbf{\#cycles} & \textbf{\#MAC} \\ \hline
CNN (CIFAR10)  & 78.9              & 3C-1D             & 150.0M            & 12.3M             \\ \hline
LeNet5         & 97.6              & 2C-3D             & 10.4M             & 423K             \\ \hline
Mcunet-vww1    & 88.9              & 1C-15R-1D         & 181.8M            & 12M             \\ \hline
MobileNetV1    & 70.4              & 14C-1D            & 5.8B              & 573M             \\ \hline
\end{tabular}
\begin{tablenotes}\small
\item[] $^1$Network's topology in terms of convolutional (C), dense (D) and residuals (R) layers
\vspace{-2ex}
\end{tablenotes}
\end{threeparttable}
\label{tab:baseline}
\end{table}

\begin{figure}
    \centering
    \includegraphics{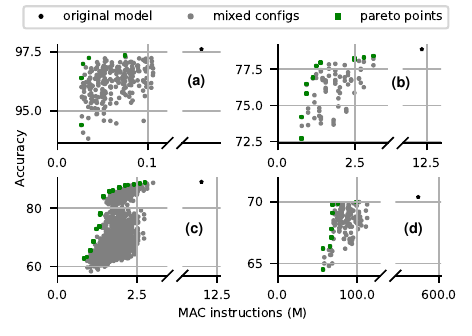}
    \caption{Accuracy - MAC instructions Pareto Space obtained from our mixed-precision configuration exploration for \blue{LeNet(a), CNN\_CIFAR10(b), MCUNet(c) and MobileNetV1(d)}.}
    \label{fig:dse}\vspace{-2ex}
\end{figure}

\begin{figure}
    \centering
    \includegraphics[width=\columnwidth]{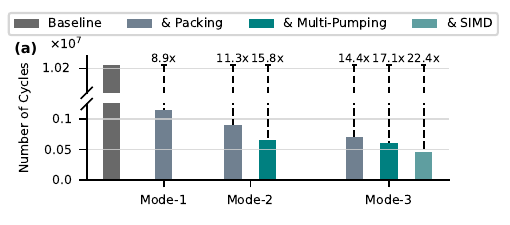} \\
    \includegraphics[width=\columnwidth]{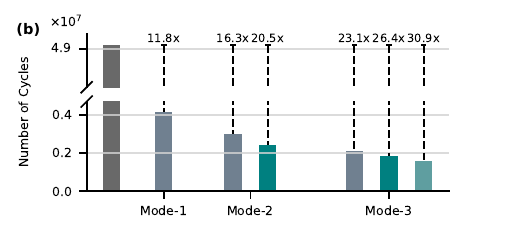}
    \caption{Processing time in cycles of a dense layer (a) and a convolution layer (b) when each $Mode$ is utilized. The distinct impact of each optimization per $Mode$ is shown.}
    \label{fig:modeeval}\vspace{-2ex}
\end{figure}

\begin{figure*}
    \centering
    \includegraphics{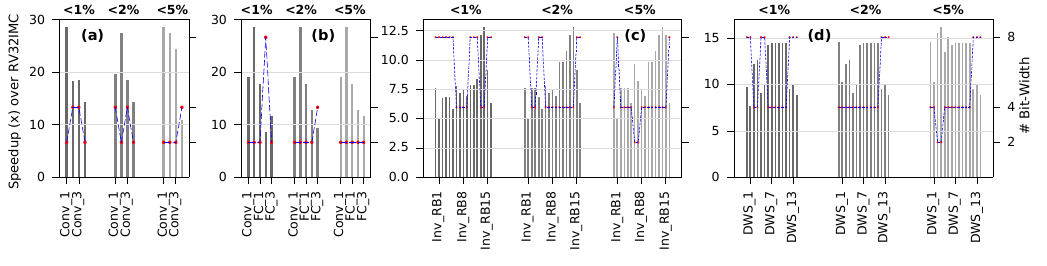}
    \caption{Speedup gains compared to the baseline Ibex RV32IMC for 4 different benchmarks, i.e., CIFAR10 CNN (a), LeNet (b), MCUNet (c) and MobileNetV1 (d), are shown. For each benchmark three different mixed-precision models have been selected, utilizing three different accuracy loss threshold (1\%, 2\%, 5\%). The bit-width of weight operands is depicted in the right Y-axis.}
    \label{fig:speedups}\vspace{-2ex}
\end{figure*}

In this section we assess the efficiency of our proposed framework by performing an in-depth evaluation over four state-of-the-art DNNs trained on four, both simple and challenging, image classification datasets (see Table~\ref{tab:baseline}).
For our analysis we consider: LeNet5, a CNN with 3 convolutional layers~\cite{CMSIS-NN}, MCUNet~\cite{MCUNet} and MobileNetV1~\cite{mbnet}, trained on MNIST, CIFAR10, Visual Wake Words~\cite{VWW} and ImageNet dataset, respectively.
Our mixed-precision post-training quantization was performed using PyTorch libraries \orange{by selecting 10\% of the training dataset, while the fine-tuning process with quantization-aware training was executed using a portion of the same dataset} and for 18 epochs, on average, requiring less than 6 hours each, on an NVDIA T4 GPU with 16GB RAM. 

For hardware results, we evaluate both ASIC and FPGA synthesis flow for the microarchitecture of the open-source Ibex core~\cite{ibexproc}, described in system verilog.
\orange{For the FPGA workflow we consider a Virtex-7 FPGA as a proof-of-concept platform, with the main core operating at 50MHz and the custom functional unit running at 100 MHz} and used Vivado 2023.1 for synthesis and power results. 
Register file is implemented using RAM32M primitives.
Cycle-accurate simulations for both performance metrics and top-1 classification accuracy, are performed using Verilator~\cite{verilator}, which reads Ibex performance counters for precise report of total cycles.
\orange{On the other hand, for the ASIC workflow the processor is designed in System Verilog RTL, synthesized with a dual clock configuration, operating at 250MHz and 500 MHz for the main core and the functional unit respectively} and is mapped to the open-source 7nm ASAP7 library~\cite{asap7}, while Synopsys Design Compiler with the $compile\_ultra$ command is used.
Register file is implemented using latches.
For power analysis, the VCD file extracted from the Verilator along with the ddc of the synthesized design were used to obtain the switching activity and the respective power analysis.

\subsection{Evaluation of our framework}\label{sec:evalframe}

\begin{table}[t]
\setlength\tabcolsep{1pt}
\renewcommand{\arraystretch}{1.1}
\small
\caption{Performance comparison of our modified Ibex processor with the baseline. Models with less than 1\% accuracy loss were considered.}
\begin{threeparttable}
\begin{tabular}{|c||cc||cc|}
\hline
\textbf{Platform} &
  \multicolumn{2}{c||}{\textbf{FPGA}} &
  \multicolumn{2}{c|}{\textbf{ASAP 7nm}} \\ \hline
\textbf{} &
  \multicolumn{1}{c|}{\textbf{\begin{tabular}[c]{@{}c@{}}Baseline\\ Ibex\end{tabular}}} &
  \textbf{\begin{tabular}[c]{@{}c@{}}Modifed\\ Ibex\end{tabular}} &
  \multicolumn{1}{c|}{\textbf{\begin{tabular}[c]{@{}c@{}}Baseline\\ Ibex\end{tabular}}} &
  \textbf{\begin{tabular}[c]{@{}c@{}}Modifed\\ Ibex\end{tabular}} \\ \hline
\textbf{Clock Freq.} &
  \multicolumn{1}{c|}{50MHz} &
  \begin{tabular}[c]{@{}c@{}}50/\tnote{1}\\ 100MHz\end{tabular} &
  \multicolumn{1}{c|}{250MHz} &
  \begin{tabular}[c]{@{}c@{}}250/\tnote{1}\\ 500MHz\end{tabular} \\ \hline
\textbf{Precision} &
  \multicolumn{1}{c|}{32-bit} &
  2-/4-/8-bit &
  \multicolumn{1}{c|}{32-bit} &
  2-/4-/8-bit \\ \hline
\textbf{Power} &
  \multicolumn{1}{c|}{256mW\tnote{2}} &
  261mW\tnote{2} &
  \multicolumn{1}{c|}{0.43mW} &
  0.58mW \\ \hline
\textbf{\begin{tabular}[c]{@{}c@{}}Resources/\\ Area\end{tabular}} &
  \multicolumn{1}{c|}{\begin{tabular}[c]{@{}c@{}}5.5K FF\\ 5.1 LUTs\\ 4 DSPs\end{tabular}} &
  \begin{tabular}[c]{@{}c@{}}7.4K FF\\ 6.4 LUTs\\ 4 DSPs\end{tabular} &
  \multicolumn{1}{c|}{0.028 $mm^2$} &
  0.038 $mm^2$ \\ \hline
\textbf{\begin{tabular}[c]{@{}c@{}}Energy\\ Efficiency\\ (GOP/s/W)\end{tabular}} &
  \multicolumn{1}{l|}{\begin{tabular}[c]{@{}l@{}}LeNet:      0.016\\ CNN:        0.032\\ MCUNet:  0.026\\ MobileV1: 0.039\end{tabular}} &
  \begin{tabular}[c]{@{}c@{}}0.34\\ 0.65\\ 0.18\\ 0.45\end{tabular} &
  \multicolumn{1}{l|}{\begin{tabular}[c]{@{}l@{}}LeNet:       47.1\\ CNN:         95.1\\ MCUNet:  76.8\\ MobileV1: 114.9\end{tabular}} &
  \begin{tabular}[c]{@{}c@{}}758.0\\ 1471.8\\ 415.61\\ 1014.9\end{tabular} \\ \hline
\end{tabular}
\begin{tablenotes}\small
\item[] $^1$Basic core's clock frequency / Multi-pumped units' clock frequency
\item[] $^2$28\% leakage power
\vspace{-2ex}
\end{tablenotes}
\end{threeparttable}
\label{tab:ibex}\vspace{-2ex}
\end{table}

\begin{table*}[h!]
\setlength\tabcolsep{4pt}
\renewcommand{\arraystretch}{1.1}
\caption{Comparison with state-of-the-art solutions. Performance metrics ordered according to supported precision configurations, while for our energy efficiency only <1\% designs were considered.}
\begin{threeparttable}
\begin{tabular}{cc|ccccccc}
\hline
\textbf{} &
  \textbf{} &
  \multicolumn{1}{c|}{\textbf{TC'24\cite{wang2024optimizing}}} &
  \multicolumn{1}{c|}{\textbf{HPCA'23\cite{Mix-GEMM}}} &
  \multicolumn{1}{c|}{\textbf{ISVLSI'20\cite{ottavi2020mixed}}} &
  \multicolumn{1}{c|}{\textbf{JSSC'18\cite{lee2018unpu}}} &
  \multicolumn{1}{c|}{\textbf{TCAD'20\cite{tcad20}}} &
  \multicolumn{1}{c|}{\textbf{DATE'20\cite{XpulpNN}}} &
  \multicolumn{1}{c}{\textbf{Ours}} \\ \hline
\multicolumn{1}{c|}{\multirow{4}{*}{\textbf{\rotatebox{90}{\begin{tabular}[c]{@{}c@{}}Archite-\\ cture\end{tabular}}}}} &
  \textbf{Platform} &
  90nm &
  22nm &
  22nm &
  65nm &
  65nm &
  22nm &
  7nm \\ \cline{2-9} 
\multicolumn{1}{c|}{} &
  \textbf{Precision} &
  32 bit &
  2-8 bit &
  2/4/8 bit &
  \multicolumn{1}{c}{1-16 bit} &
  16 bit &
  \multicolumn{1}{c}{2/4/8 bit} &
  \multicolumn{1}{c}{2/4/8 bit} \\ \cline{2-9} 
\multicolumn{1}{c|}{} &
  \textbf{Clk Freq.} &
  100MHz &
  1.2GHz &
  250MHz &
  2500MHz &
  200MHz &
  600MHz &
  250MHz \\ \cline{2-9} 
\multicolumn{1}{c|}{} &
  \textbf{Area/Power} &
  \begin{tabular}[c]{@{}c@{}}6.44$mm^2$/ \\ 5.8mW\end{tabular} &
  \begin{tabular}[c]{@{}c@{}}0.014$mm^2$/\\ 9.9mW\tnote{3}\end{tabular} &
  \begin{tabular}[c]{@{}c@{}}0.002 $mm^2$/\\ 5.5mW\tnote{3}\end{tabular} &
  \begin{tabular}[c]{@{}l@{}}16 $mm^2$/\\ 288mW\end{tabular} &
  \begin{tabular}[c]{@{}l@{}}11.47$mm^2$/\\ 805mW\end{tabular} &
  \begin{tabular}[c]{@{}l@{}}0.04 $mm^2$/\\ 43.5mW\tnote{3}\end{tabular} &
  \begin{tabular}[c]{@{}l@{}}0.038$mm^2$/\\ 0.58mW\end{tabular} \\ \hline \hline
\multicolumn{1}{c|}{\multirow{2}{*}{\rotatebox{90}{\textbf{\begin{tabular}[c]{@{}c@{}}Perform-\\ ance\end{tabular}}}}} &  
  \textbf{GOPs} &
  0.23 &
  11.9 &
  3.3 &
  514.2 &
  288.0 &
  47.9 &
  0.24-0.85 \\ \cline{2-9} 
\multicolumn{1}{c|}{} &
  \textbf{\begin{tabular}[c]{@{}c@{}}Energy Eff.\\ (GOPs/W)\end{tabular}} &
  38.8\tnote{1} &
  500-1166\tnote{2} &
  200-600\tnote{1} &
  1750\tnote{2} &
  357.8\tnote{2} &
  700-1100\tnote{1} &
  415-1470 \\ \hline
\end{tabular}
\begin{tablenotes}\small
\item[] $^1$Throughput and energy efficiency were calculated based on a typical convolution layer $^3$Area includes only extended units
\item[] $^2$Throughput and energy efficiency were calculated based on the average values of examined DNNs 
\vspace{-2ex}
\end{tablenotes}
\end{threeparttable}
\label{tab:soa}
\end{table*}

As mentioned, our framework starts with a design space exploration to find different mixed-precision configurations for each input benchmark.
Fig.~\ref{fig:dse} presents the Pareto space between accuracy and required number of \blue{MAC instructions} for all the DNNs examined.
In Fig.~\ref{fig:dse}, the black star represents the pre-trained baseline model.
The gray circles are the quantized models in mixed precision among their layers, while the green squares form the Pareto front.
Although at this stage a software exploration is performed only, the reduction in the amount of MAC instructions is obtained based on the specific bit-width per layer and the degree of parallelization that can be achieved due to the introduced packing, multi-pumping and SIMD.
Note that the number of the explored models depends on the number of layers.
In total, to generate Fig.~\ref{fig:dse}, we evaluated more than 3500 quantized models.
It is observed that different precisions in layers ranging between 2,4 and 8 bits, can achieve accuracies comparable to or the same with the full-precision counterparts.
For all benchmarks more than 86\% reduction in the number of \blue{MAC instructions} can be achieved in their mixed quantized derivatives for less than 1\% accuracy loss, while this number increases to 93\% when up to 5\% accuracy loss is applicable.

The impact of each distinct optimization technique (different $Modes$) that is applied within a processor execution is evaluated in Fig.~\ref{fig:modeeval}.  
To generate Fig.~\ref{fig:modeeval} we consider as an example only one distinct dense layer (Fig.~\ref{fig:modeeval}a), i.e., the final layer of the MobileNetV1, and a convolution layer (Fig.~\ref{fig:modeeval}b), i.e., the second layer of the CIFAR10 CNN, and evaluate the relative speedups of the standalone $Mode-1,-2,-3$.
In other words, we report the gains of each technique as if it was solely applied throughout the whole layer and examined for all the three different bit-widths.
It is observed, that $Mode-1$ due to the high parallelization of 8-bit operands and minimization of loads and stores, can achieve, on average, 9.9x speedup compared to the baseline RV32IMC, and 17.8x when 2-bit weights are utilized.
On the other hand, when the standalone multi-pumping technique is applied ($Mode-2$), it can bring an extra 16\% average speedup to the total execution cycles per layer, for both 4-bit and 2-bit cases.
Finally, when 2-bit weights are considered ($Mode-3$), our soft SIMD approach obtains an extra 13\% speedup, on average, having a significant reduction in total cycles of up to 30.9x.

As mentioned earlier, the sensitivity and error resilience of each layer varies, and hence top-1 models' accuracy, is tightly dependent on the selected precision for each layer. 
Subsequently, in order to put the delivered trade-offs between total execution time of DNN inference and top-1 accuracy into perspective, we generate Fig.~\ref{fig:speedups}.
This figure considers three conservative accuracy loss thresholds (set by the user) and extracts three optimal mixed-precision configurations obtained from the DSE.
The bit-width configurations of the weight operands per layer are also shown on the right Y-axis.
Then, the speedup for all layers derived from Verilator is reported.
The key observations regarding the mixed precision selection are two: 1) the less complex models like LeNet and CIFAR10 CNN allow aggressive quantization down to 2 bits in most of their layers with minimal accuracy degradation (<$1\%$) and thus able to achieve the maximum possible performance gains, and 2) although the more challenging MobileNetV1 and MCUNet models barely utilize 2-bit weights scenarios (only for >5\% accuracy loss constraints), they manage to compress most of their layers to 4-bit with small impact on their accuracy (<2\%).
Overall and on average for all layers, our proposed ISA extensions can achieve from 13.1x up to 17.8x speedup for 1\% up to 5\% accuracy degradation, respectively.
It is also noteworthy that MCUNet presents less significant gains compared to other benchmarks due to the high amount of depthwise convolutions.
The latter do not enable the same degree of input reuse as in standard point-wise convolutions, while also they differ in the overheads (e.g., branch instructions) they introduce.

\textbf{FPGA and ASIC Designs comparisons:}
In Table~\ref{tab:ibex} we compare the performance of the proposed modified Ibex processor with the baseline one on both FPGA and ASIC platforms. 
On one hand, our proposed modified processor, operating on an FPGA at a clock frequency of only 50MHz \orange{(and 100 MHz for the multi-pumped circuit)}, primarily due to the costly RAM32M-based register file, achieves, averagely for all benchmarks, energy efficiency improvement of $15\times$ over the baseline processor. 
Despite experiencing approximately a 25\% increase in resource utilization for LUTs and FFs, along with a 2\% power overhead, it is evident that our extended RISC-V ISA, tailored to our hardware optimizations, can effectively deliver substantial energy efficiency gains, i.e., $14.3\times$ higher than baseline Ibex.
Alternatively, notable gains are also evident for our ASIC design.
In particular, when analyzing models with less than a 1\% accuracy loss, we observe an 11$\times$ improvement in energy efficiency, with only a 25.8\% increase in power and a 26.3\% increase in area.
Note also that power the consumption of a latch-based register file implementation\footnote{We were not able to have access to a memory compiler for ASAP 7nm, thus the Register File has been implemented with conventional latch-based standard cells.} has been factored in as well, with potentially even higher gains anticipated when SRAM is integrated instead ~\cite{1522758}.

\subsection{Comparison against state-of-the-art}\label{sec:soacomp}


In this section, we compare our modified RISC-V processor with the most relevant state-of-the-art solutions. We note that our solution utilizes a limited standard cell technology library provided by the ASAP open academic 7nm PDK, while other solutions are mapped on industrial strength technology libraries  
More specifically, we analyze hardware-software co-designed architectures computing DNNs on CPU architectures, adopting ISA extensions and custom functional units.
A detailed comparison is shown in Table~\ref{tab:soa}.
To compare the performance of our modified RISC-V processor equipped with custom ISA extension with respect to state-of-the-art, we report the peak performance presented in each work.
For an as fair as possible comparison, either a typical convolution layer, or the average peak performance for all the examined DNNs was considered.
In other works, accuracy degradation of the corresponding peak performance of related works, where in some cases could be high~\cite{Mix-GEMM}, is not examined at this point.
On the other hand, our evaluation presents an energy efficiency range that corresponds to less than 1\% (see Table~\ref{tab:ibex}) up to 5\% accuracy loss.

As shown in Table~\ref{tab:soa}, our average lowest performance (915 GOPs/W) surpasses the corresponding value of MIX-GEMM~\cite{Mix-GEMM} for less than 1\% accuracy degradation, while compared to~\cite{ottavi2020mixed}, our performance ranges from 0.7$\times$ up to 3.2$\times$ faster.
When considering MobileNetV1 specifically, Mix-GEMM achieves energy efficiency ranging from 500 to 1000 GOPs/W, while our approach surpasses this performance, with respective numbers ranging from 1015 to 1149 GOPs/W. 
On the other hand, XPulpNN~\cite{XpulpNN}, which utilizes SIMD units supporting from 4 8-bit to 16 2-bit MAC per cycle, reports comparable performance when processing a typical convolution layer. However, its efficiency stands 25\% lower than that of our processor when compared to our CNN model.
Additionally, although UNPU's decoupled accelerator~\cite{lee2018unpu} showcases high performance, its throughput is hindered by its limited flexibility, since the intricate software stack required by such accelerators, usually demands specific offloading mechanisms handled at either the hardware or software level.
Finally, note that only designs with 1\% accuracy loss constraints were evaluated for Table~\ref{tab:soa}.
However, when prioritizing efficiency over top-1 accuracy (e.g., up to 5\% degradation), our processor achieves an average of 1.07 TOps/W, with a peak performance of 1.9 TOPs/W on the MobileNetV1 Imagenet model.

\section{conclusion}

In this work, we embrace the potential of mixed-precision neural networks, trying to overcome the limitations of current CPU architectures that hinder efficient precision management.
Our approach involves enhancing a RISC-V CPU core with lightweight hardware modifications and an end-to-end automated framework for generating and evaluating mixed-precision neural networks. 
We introduce three novel instructions to the RISC-V ISA, enabling various operational modes for optimized computations that leverage operand packing, multi-pumping and a soft SIMD technique.
Our advancements through a RISC-V based processor demonstrate that not only accuracy can be preserved, but also remarkable gains in energy efficiency and performance can be achieved.
The latter, represents a substantial leap forward in energy-efficient computing for resource-constrained devices.

{\small
\begin{acks}
This work is partially supported by EU Horizon research and
innovation programme, under project CONVOLVE, grant agreement
No 101070374.
\end{acks}
}
\bibliographystyle{IEEEtran}
\bibliography{IEEEtran}

\begin{thebibliography}{10}
\providecommand{\url}[1]{#1}
\def\UrlFont{\rmfamily}
\providecommand{\newblock}{\relax}
\providecommand{\bibinfo}[2]{#2}
\providecommand\BIBentrySTDinterwordspacing{\spaceskip=0pt\relax}
\providecommand\BIBentryALTinterwordstretchfactor{4}
\providecommand\BIBentryALTinterwordspacing{\spaceskip=\fontdimen2\font plus
\BIBentryALTinterwordstretchfactor\fontdimen3\font minus \fontdimen4\font\relax}
\providecommand\BIBforeignlanguage[2]{{%
\expandafter\ifx\csname l@#1\endcsname\relax
\typeout{** WARNING: IEEEtran.bst: No hyphenation pattern has been}%
\typeout{** loaded for the language `#1'. Using the pattern for}%
\typeout{** the default language instead.}%
\else
\language=\csname l@#1\endcsname
\fi
#2}}

\bibitem{jouppi2017datacenter}
N.~P. Jouppi, C.~Young, N.~Patil, D.~Patterson, G.~Agrawal, R.~Bajwa, S.~Bates, S.~Bhatia, N.~Boden, A.~Borchers, \emph{et~al.}, ``In-datacenter performance analysis of a tensor processing unit,'' in \emph{Proceedings of the 44th annual international symposium on computer architecture}, 2017, pp. 1--12.

\bibitem{Pulp-NN}
A.~Garofalo, M.~Rusci, F.~Conti, D.~Rossi, and L.~Benini, ``Pulp-nn: A computing library for quantized neural network inference at the edge on risc-v based parallel ultra low power clusters,'' in \emph{2019 26th IEEE International Conference on Electronics, Circuits and Systems (ICECS)}, 2019, pp. 33--36.

\bibitem{Mix-GEMM}
E.~Reggiani, A.~Pappalardo, M.~Doblas, M.~Moreto, M.~Olivieri, O.~S. Unsal, and A.~Cristal, ``Mix-gemm: An efficient hw-sw architecture for mixed-precision quantized deep neural networks inference on edge devices,'' in \emph{2023 IEEE International Symposium on High-Performance Computer Architecture (HPCA)}, 2023, pp. 1085--1098.

\bibitem{csurarme}
G.~Armeniakos, G.~Zervakis, D.~Soudris, and J.~Henkel, ``Hardware approximate techniques for deep neural network accelerators: A survey,'' \emph{ACM Comput. Surv.}, vol.~55, no.~4, nov 2022.

\bibitem{XpulpNN}
A.~Garofalo, G.~Tagliavini, F.~Conti, D.~Rossi, and L.~Benini, ``Xpulpnn: Accelerating quantized neural networks on risc-v processors through isa extensions,'' in \emph{2020 Design, Automation \& Test in Europe Conference \& Exhibition (DATE)}, 2020, pp. 186--191.

\bibitem{Eyeriss}
Y.-H. Chen, T.~Krishna, J.~S. Emer, and V.~Sze, ``Eyeriss: An energy-efficient reconfigurable accelerator for deep convolutional neural networks,'' \emph{IEEE Journal of Solid-State Circuits}, vol.~52, no.~1, pp. 127--138, 2017.

\bibitem{ibexproc}
www.lowrisc.org, ``Ibex risc-v core,'' \url{https://github.com/lowRISC/ibex}.

\bibitem{10069694}
C.-J. Johnsen, T.~D. Matteis, T.~Ben-Nun, J.~d. Fine~Licht, and T.~Hoefler, ``Temporal vectorization: A compiler approach to automatic multi-pumping,'' in \emph{2022 IEEE/ACM International Conference On Computer Aided Design (ICCAD)}, 2022, pp. 1--9.

\bibitem{multipump:2017}
R.~Zhao, T.~Todman, W.~Luk, and X.~Niu, ``Deeppump: Multi-pumping deep neural networks,'' in \emph{2017 IEEE 28th International Conference on Application-specific Systems, Architectures and Processors (ASAP)}, 2017, pp. 206--206.

\bibitem{ottavi2020mixed}
G.~Ottavi, A.~Garofalo, G.~Tagliavini, F.~Conti, L.~Benini, and D.~Rossi, ``A mixed-precision risc-v processor for extreme-edge dnn inference,'' in \emph{2020 IEEE Computer Society Annual Symposium on VLSI (ISVLSI)}.\hskip 1em plus 0.5em minus 0.4em\relax IEEE, 2020, pp. 512--517.

\bibitem{XpulpNNV2}
\BIBentryALTinterwordspacing
N.~Bruschi, A.~Garofalo, F.~Conti, G.~Tagliavini, and D.~Rossi, ``Enabling mixed-precision quantized neural networks in extreme-edge devices,'' in \emph{Proceedings of the 17th ACM International Conference on Computing Frontiers}, ser. CF '20.\hskip 1em plus 0.5em minus 0.4em\relax New York, NY, USA: Association for Computing Machinery, 2020, p. 217–220. [Online]. Available: \url{https://doi.org/10.1145/3387902.3394038}
\BIBentrySTDinterwordspacing

\bibitem{lee2018unpu}
J.~Lee, C.~Kim, S.~Kang, D.~Shin, S.~Kim, and H.-J. Yoo, ``Unpu: An energy-efficient deep neural network accelerator with fully variable weight bit precision,'' \emph{IEEE Journal of Solid-State Circuits}, vol.~54, no.~1, pp. 173--185, 2018.

\bibitem{tcad20}
W.~Xu, Z.~Zhang, X.~You, and C.~Zhang, ``Reconfigurable and low-complexity accelerator for convolutional and generative networks over finite fields,'' \emph{IEEE Transactions on Computer-Aided Design of Integrated Circuits and Systems}, vol.~39, no.~12, pp. 4894--4907, 2020.

\bibitem{wang2024optimizing}
S.~Wang, X.~Wang, Z.~Xu, B.~Chen, C.~Feng, Q.~Wang, and T.~T. Ye, ``Optimizing cnn computation using risc-v custom instruction sets for edge platforms,'' \emph{IEEE Transactions on Computers}, 2024.

\bibitem{FILMQNNEF}
\BIBentryALTinterwordspacing
M.~Sun, Z.~Li, A.~Lu, Y.~Li, S.-E. Chang, X.~Ma, X.~Lin, and Z.~Fang, ``Film-qnn: Efficient fpga acceleration of deep neural networks with intra-layer, mixed-precision quantization,'' in \emph{Proceedings of the 2022 ACM/SIGDA International Symposium on Field-Programmable Gate Arrays}, ser. FPGA '22.\hskip 1em plus 0.5em minus 0.4em\relax New York, NY, USA: Association for Computing Machinery, 2022, p. 134–145. [Online]. Available: \url{https://doi.org/10.1145/3490422.3502364}
\BIBentrySTDinterwordspacing

\bibitem{SySMOL}
C.~Zhou, V.~Richard, P.~Savarese, Z.~Hassman, M.~Maire, M.~DiBrino, and Y.~Li, ``Sysmol: A hardware-software co-design framework for ultra-low and fine-grained mixed-precision neural networks,'' 2023.

\bibitem{zervtc}
K.~Balaskas, A.~Karatzas, C.~Sad, K.~Siozios, I.~Anagnostopoulos, G.~Zervakis, and J.~Henkel, ``Hardware-aware dnn compression via diverse pruning and mixed-precision quantization,'' \emph{IEEE Transactions on Emerging Topics in Computing}, pp. 1--14, 2024.

\bibitem{CMSIS-NN}
L.~Lai, N.~Suda, and V.~Chandra, ``Cmsis-nn: Efficient neural network kernels for arm cortex-m cpus,'' \emph{arXiv preprint arXiv:1801.06601}, 2018.

\bibitem{jacob2022gemmlowp}
B.~Jacob and P.~Warden, ``gemmlowp: A small self-contained low-precision gemm library,'' \url{https://github.com/google/gemmlowp}, 2022.

\bibitem{STM-XCUBE-AI}
{STMicroelectronics}, ``{X-CUBE-AI - Artificial Intelligence Expansion Package},'' \url{https://www.st.com/en/embedded-software/x-cube-ai.html}, 2023.

\bibitem{Dory}
A.~Burrello, A.~Garofalo, N.~Bruschi, G.~Tagliavini, D.~Rossi, and F.~Conti, ``Dory: Automatic end-to-end deployment of real-world dnns on low-cost iot mcus,'' \emph{IEEE Transactions on Computers}, vol.~70, no.~8, pp. 1253--1268, 2021.

\bibitem{MCUNet}
J.~Lin, W.-M. Chen, Y.~Lin, C.~Gan, S.~Han, \emph{et~al.}, ``Mcunet: Tiny deep learning on iot devices,'' \emph{Advances in Neural Information Processing Systems}, vol.~33, pp. 11\,711--11\,722, 2020.

\bibitem{CMix-NN}
A.~Capotondi, M.~Rusci, M.~Fariselli, and L.~Benini, ``Cmix-nn: Mixed low-precision cnn library for memory-constrained edge devices,'' \emph{IEEE Transactions on Circuits and Systems II: Express Briefs}, vol.~67, no.~5, pp. 871--875, 2020.

\bibitem{YodaNN}
R.~Andri, L.~Cavigelli, D.~Rossi, and L.~Benini, ``Yodann: An architecture for ultralow power binary-weight cnn acceleration,'' \emph{IEEE Transactions on Computer-Aided Design of Integrated Circuits and Systems}, vol.~37, no.~1, pp. 48--60, 2018.

\bibitem{Barvinn}
\BIBentryALTinterwordspacing
M.~Askarihemmat, S.~Wagner, O.~Bilaniuk, Y.~Hariri, Y.~Savaria, and J.-P. David, ``Barvinn: Arbitrary precision dnn accelerator controlled by a risc-v cpu,'' in \emph{Proceedings of the 28th Asia and South Pacific Design Automation Conference}, ser. ASPDAC '23.\hskip 1em plus 0.5em minus 0.4em\relax New York, NY, USA: Association for Computing Machinery, 2023, p. 483–489. [Online]. Available: \url{https://doi.org/10.1145/3566097.3567872}
\BIBentrySTDinterwordspacing

\bibitem{DNN_Intel_FPGA}
P.~Colangelo, N.~Nasiri, E.~Nurvitadhi, A.~Mishra, M.~Margala, and K.~Nealis, ``Exploration of low numeric precision deep learning inference using intel® fpgas,'' in \emph{2018 IEEE 26th Annual International Symposium on Field-Programmable Custom Computing Machines (FCCM)}, 2018, pp. 73--80.

\bibitem{Approximate_Computing_Survey}
V.~Leon, M.~A. Hanif, G.~Armeniakos, X.~Jiao, M.~Shafique, K.~Pekmestzi, and D.~Soudris, ``Approximate computing survey, part i: Terminology and software \& hardware approximation techniques,'' 2023.

\bibitem{spike}
B.~Keller, ``Risc-v, spike, and the rocket core,'' \emph{Berkeley Architecture Group}, 2013.

\bibitem{verilator}
W.~Snyder, P.~Wasson, D.~Galbi, and {et al}, ``{Verilator},'' \url{https://github.com/verilator/verilator}.

\bibitem{mbnet}
A.~G. Howard, M.~Zhu, B.~Chen, D.~Kalenichenko, W.~Wang, T.~Weyand, M.~Andreetto, and H.~Adam, ``Mobilenets: Efficient convolutional neural networks for mobile vision applications,'' \emph{arXiv preprint arXiv:1704.04861}, 2017.

\bibitem{VWW}
A.~Chowdhery, P.~Warden, J.~Shlens, A.~Howard, and R.~Rhodes, ``Visual wake words dataset,'' \emph{arXiv preprint arXiv:1906.05721}, 2019.

\bibitem{asap7}
L.~T. Clark, V.~Vashishtha, L.~Shifren, A.~Gujja, S.~Sinha, B.~Cline, C.~Ramamurthy, and G.~Yeric, ``Asap7: A 7-nm finfet predictive process design kit,'' \emph{Microelectronics Journal}, vol.~53, pp. 105--115, 2016.

\bibitem{1522758}
Y.~Li, M.~Hempstead, P.~Mauro, D.~Brooks, Z.~Hu, and K.~Skadron, ``Power and thermal effects of sram vs. latch-mux design styles and clock gating choices,'' in \emph{ISLPED '05. Proceedings of the 2005 International Symposium on Low Power Electronics and Design, 2005.}, 2005, pp. 173--178.

\end{thebibliography}

\end{document}